\documentclass[10pt,prd,aps,twocolumn,nofootinbib,superscriptaddress,eqsecnum,floatfix,preprintnumbers,amsmath,amssymb,longbibliography]{revtex4-2}

\makeatletter
\renewcommand{\p@subsection}{}
\renewcommand{\p@subsubsection}{}
\makeatother

\usepackage{graphicx}
\usepackage{epsfig}
\usepackage[dvipsnames]{xcolor}
\usepackage[utf8]{inputenc}
\usepackage{stmaryrd}
\usepackage{mathrsfs}
\usepackage{mathalfa}
\usepackage{accents}
\usepackage{enumitem}
\usepackage[normalem]{ulem}

\usepackage[colorlinks=true,
            citecolor=green,
            linkcolor=red,
            filecolor=cyan,
            urlcolor=magenta,
            backref=false]{hyperref}

\DeclareMathOperator\arctanh{arctanh}

\newcommand{\dd}{\mathrm{d}}

\newcommand{\be}{\begin{equation}}
\newcommand{\ee}{\end{equation}}
\newcommand{\bea}{\begin{eqnarray}}
\newcommand{\eea}{\end{eqnarray}}
\newcommand{\ba}{\begin{equation}\begin{aligned}}
\newcommand{\ea}{\end{aligned}\end{equation}}
\newcommand{\beg}{\begin{gather*}}
\newcommand{\eng}{\end{gather*}}
\newcommand{\hh}{,\hspace{0.5cm}}

\newcommand{\n}[1]{\label{#1}}

\newcommand{\CAL}{\mathcal}

\newcommand{\bs}{\begin{split}}
\newcommand{\es}{\end{split}}

\usepackage{graphicx} 

\begin{document}

\title{QTG MINIMONS}

\author{Valeri P. Frolov}%
\email[]{vfrolov@ualberta.ca}
\affiliation{Theoretical Physics Institute, Department of Physics,
University of Alberta,\\
Edmonton, Alberta, T6G 2E1, Canada
}

\begin{abstract}
We study static, spherically symmetric, regular horizonless solutions
of quasitopological gravity (QTG), which we call QTG minimons. These
configurations form the small-mass branch below the threshold for
regular black holes. We include vacuum polarization by coupling the
spherically reduced QTG action to the Polyakov action for
two-dimensional conformal matter. For an arbitrary characteristic
function $h(p)$, we derive a constrained first-order system in the
static, asymptotically empty state. We obtain central expansions in
$D\ge4$ and analyze four-dimensional solutions in the Hayward-type and
Born--Infeld-type models. Both models retain a de Sitter-like core.
Along the near-critical branch, the asymptotic mass decreases with the
quantum coupling, while the central blueshift can become large even
for a small coupling. The numerical results indicate that vacuum
polarization is a nonuniform perturbation near the critical limit.
Assuming a smooth constant-radius limit in proper distance, we derive
a candidate $\mathrm{AdS}_2\times S^{D-2}$ throat geometry for the
limiting configurations, which we call QTG extremons.
\end{abstract}
\hfill    Alberta Thy 8-26

\date{November 2026}

\maketitle

Regular black hole solutions in quasitopological gravity (QTG) have
recently attracted considerable attention
\cite{QT_BH,Bueno:2025zaj,Myers:2010ru,rbh_pfkz,Bueno:2024zsx,DiFilippo:2024mwm,Bueno:2024eig,Bueno:2025tli,Bueno:2025gjg,PinedoSoto:2025hel,Bueno:2026dln,Borissova:2026wmn,Colleaux:2026hat,Borissova:2026prd}.
In  these models, the Einstein--Hilbert
action is supplemented by an infinite series of scalar polynomials in
the curvature. These polynomials are chosen so that, in spherical
symmetry, the field equations contain no derivatives of the metric
higher than second order. The spherically reduced action is
characterized by a function $h(p)$ of the primary curvature invariant
$p$ (see, e.g., \cite{Bueno:2020gq} and references therein). A broad
class of such functions gives rise to regular black hole solutions.

These models contain a characteristic length $\ell$, which sets the
scale at which higher-curvature corrections become important. Since
curvature has dimension $[\mathrm{length}]^{-2}$, the relative strengths
of successive curvature terms involve appropriate powers of $\ell$.
When the spacetime curvature reaches values of order $\ell^{-2}$,
these terms can substantially modify the Einstein equations.

Although the analysis of QTG is restricted to spherical symmetry, this sector allows us to address a wide range of problems, including black hole formation and evaporation
\cite{Bueno:2024zsx,Bueno:2024eig,Bueno:2025gjg,Borissova:2026prd},
spherical accretion \cite{Frolov:2026tft}, regular black holes in
de Sitter and anti-de Sitter spacetimes, charged generalizations
\cite{BazrafshanGhanaatian2013,PinedoSoto:2026hfm,Frolov:2026tft},
and the mass-inflation problem
\cite{Poisson:1990eh,Ori:1991zz,Bonanno:2020fgp,Carballo-Rubio:2021bpr,Carballo-Rubio:2022kad,Carballo-Rubio:2024dca,DiFilippo:2024mwm,Frolov:2026rcm,Borissova:2026kwu,DiFilippo:2026jpv}.
A modified double-copy approach also extends the well-known
double-copy formalism of general relativity to QTG black hole
solutions \cite{Frolov:2025ddw,Frolov:2026byy,Frolov:2026dcr}.

For these applications, the spherically reduced QTG action can be
expressed as a two-dimensional dilaton gravity action in the $(t,r)$
sector \cite{Bueno:2024zsx}. The original construction was developed
for $D\geq5$. Its reduced dilaton formulation can also be used to
define a four-dimensional, spherically symmetric effective model.

The static, asymptotically flat regular black hole solutions considered
here possess two horizons: an outer event horizon at $r=r_+$ and an
inner horizon at $r=r_-$. For large black hole masses, the inner
horizon lies close to the central core, with $r_-\gtrsim\ell$. As the
mass decreases, the outer horizon moves inward and the inner horizon
moves outward. At a critical value of the mass parameter,
$\mu=\mu_*$, the two horizons merge into a degenerate horizon,
forming an extremal black hole. In dimensionful variables, $\mu_*$
is of order $\ell^{D-3}$ and hence of order $\ell$ in four dimensions.
This critical value defines a mass gap for the black hole branch.
It also suggests a threshold for the production of such regular black
holes in particle collisions, although the formation process itself
requires a dynamical analysis (see e.g. \cite{Frolov:2015bta,Frolov:2015usa}).

Below the critical mass, the solutions are horizonless. In this paper,
we study these regular subcritical configurations, which we call
\emph{QTG minimons}. This terminology distinguishes them from earlier
uses of ``minimon'' for minimum-mass particles
\cite{Markov1987MaximonMinimon,Vipindas2019MaximonsMinimons} and extreme
sub-Planckian black holes \cite{Carr2015SubPlanckian}. We refer to the
limiting configurations at the upper mass bound of the minimon branch
as \emph{extremal minimons}, or \emph{extremons}.

The gravitational field of a minimon is regular throughout spacetime,
including at the center, $r=0$. For the models considered here, the
core is de Sitter-like: the primary curvature invariant approaches
$p(0)=\ell^{-2}$, while the other curvature invariants remain finite.
We assume that no classical matter sources are present for $r>0$.

In principle, the length $\ell$ is a free parameter of the effective
model. If the higher-curvature terms originate from quantum gravity,
it is natural to expect $\ell$ to be related to the Planck length
$\ell_{\mathrm{Pl}}$. We assume that $\ell$ is sufficiently larger
than $\ell_{\mathrm{Pl}}$ for metric fluctuations to remain small
and an effective description of gravity to be applicable. The validity
of this approximation is an assumption of our analysis.

Even in the absence of classical matter, quantum fields can affect
the geometry. Although a vanishing classical field configuration may
have a vanishing stress-energy tensor, quantum fluctuations can
produce a nonzero renormalized vacuum expectation value of this
tensor. Such vacuum polarization can modify minimon solutions,
particularly when $\ell$ is not much larger than
$\ell_{\mathrm{Pl}}$.

We study this effect using the two-dimensional QTG dilaton model
coupled to the Polyakov action. This action describes the
anomaly-induced contribution of massless conformal fields to the
quantum effective action. The corresponding stress-energy tensor
also depends on the quantum state, which is specified by boundary
conditions. The Polyakov action is proportional to the central
charge $C$; for independent real massless scalar fields, $C$ equals
the number of fields. This construction provides a tractable model
of vacuum polarization, rather than the full effective action of
four-dimensional quantum matter. We show that its backreaction on
minimon solutions can become important near the critical limit,
even when the dimensionless quantum coupling is small.

The paper is organized as follows. In section~2, we introduce the
spherically reduced QTG action and its coupling to the Polyakov
action, which describes vacuum polarization effects. In section~3,
we vary the combined action and derive the field equations. In
section~4, we recast these equations as a first-order system and
specify the boundary conditions. Sections~5 and~6 are devoted to
QTG minimon solutions in the Hayward-type and Born--Infeld-type
models, respectively. In section~7, we analyze near-extremal minimon
solutions and their properties. Section~8 summarizes our results
and discusses their implications.

\section{Action}

\subsection{Spherically reduced QTG action}

We write a $D$-dimensional static, spherically symmetric metric in the
form\footnote{Here and throughout, we use units in which $c=1$.}
\begin{equation}\n{METRIC}
\begin{split}
 &ds^2=d\gamma^2+r^2d\Omega_{D-2}^2\,,\\
 &d\gamma^2=\gamma_{ab}dx^a dx^b=-N^2 f\,dt^2+\frac{dr^2}{f}
\end{split}
\end{equation}
Here $N=N(r)$, $f=f(r)$, and $d\Omega_{D-2}^2$ is the metric on the
unit $(D-2)$-sphere.
After integration over the angular coordinates and omission of boundary
terms, the action of quasitopological gravity (QTG) in this sector takes
the form \cite{QT_BH,rbh_pfkz}
\begin{equation}
\begin{aligned}
 S_{\mathrm{QTG}}&=-B\int dt\,dr\,N'H\,,\\
 H&=r^{D-1}h(p)\,,\qquad B=\frac{b}{G_D}\,,\\
 b&=\frac{(D-2)\Omega_{D-2}}{16\pi}
   =\frac{(D-2)\pi^{(D-3)/2}}
   {8\Gamma\!\left(\frac{D-1}{2}\right)}\,.
\end{aligned}
\end{equation}
Here a prime denotes differentiation with respect to $r$, $G_D$ is the
$D$-dimensional Newton constant, and
\begin{equation}
 \Omega_{D-2}=\frac{2\pi^{(D-1)/2}}
 {\Gamma\!\left(\frac{D-1}{2}\right)}
\end{equation}
is the area of the unit $(D-2)$-sphere. We use the standard gravitational
action normalization $1/(16\pi G_D)$.
For $D=2$ one has $b=1/2$.

The primary curvature invariant $p$ is defined by
\begin{equation}
 p=\frac{1-f}{r^2}\,,
\end{equation}
and the function $h=h(p)$ specifies the QTG model.

In the chosen units, with $c=1$, the $D$-dimensional Newton constant
has dimension
\begin{equation}
 [G_D]=\frac{L^{D-3}}{M}\,.
\end{equation}
Since $[h(p)]=L^{-2}$, we have $[H]=L^{D-3}$. Using $[N]=1$,
$[N']=L^{-1}$, and $[dt]=[dr]=L$, we find that the reduced action
has the correct dimension,
\begin{equation}
 [S_{\mathrm{QTG}}]
 =ML\,.
\end{equation}
Here $M$ and $L$ denote the dimensions of mass and length, respectively.

QTG models contain a fundamental length scale $\ell$. When the spacetime curvature becomes of order $\ell^{-2}$, higher-curvature terms in the effective QTG action significantly modify the solutions. It is convenient to use this parameter to make spacetime coordinates and curvature dimensionless. In what follows we use this option and write the action $S_{QTG}$ as follows
\be
\n{dim}
S_{QTG}=-\dfrac{b \ell^{D-2}}{G_D}\tilde{S}_{QTG}\, .
\ee
where $\tilde{S}_{QTG}$ is dimensionless.
We retain the notation $t$ and $r$, but henceforth understand these and other dimensional quantities to have been made dimensionless using appropriate powers of the length scale $\ell$.

\subsection{Polyakov action}
\label{subsec:polyakov-action}

Let us define
$\Box=\gamma^{ab}\nabla_a\nabla_b$.
For conformal matter with total central charge $C$, the
nonlocal Polyakov action is
\begin{equation}
 S_{\mathrm P}[g]
 =-\frac{\hbar C}{96\pi}
 \int d^2x\,\sqrt{-g}\,R\Box^{-1}R\,.
 \label{eq:polyakov-nonlocal}
\end{equation}
For $n$ independent real massless scalar fields, $C=n$.
The inverse operator requires a Green function and a prescription for
boundary conditions and zero modes:
\begin{equation}
\begin{split}
&(\Box^{-1}R)(x)=\int d^2x'\,\sqrt{-\gamma(x')}\,G(x,x')R(x'),\\
&\Box_xG(x,x')=\frac{\delta^{(2)}(x-x')}{\sqrt{-\gamma(x)}}\,.
\end{split}
\end{equation}

Introducing a dimensionless auxiliary field $\varphi$, we obtain the
local representation
\begin{align}
 S_{\mathrm P}[\gamma,\varphi]
 &=\frac{\hbar C}{96\pi}
 \int d^2x\,\sqrt{-\gamma}\,
 \left(\varphi\Box\varphi-2\varphi R\right)
 \label{eq:polyakov-local}\\
 &=-\frac{\hbar C}{96\pi}
 \int d^2x\,\sqrt{-\gamma}\,
 \left[(\nabla\varphi)^2+2\varphi R\right],
 \label{eq:polyakov-local-first}
\end{align}
where the second equality holds up to boundary terms and
$(\nabla\varphi)^2=\gamma^{ab}\nabla_a\varphi\nabla_b\varphi$.
Variation with respect to $\varphi$ yields
\begin{equation}
 \Box\varphi=R\,.
 \label{eq:polyakov-auxiliary}
\end{equation}
Substituting $\varphi=\Box^{-1}R$ into
\eqref{eq:polyakov-local} reproduces \eqref{eq:polyakov-nonlocal}.
Homogeneous solutions, $\Box\varphi_{\mathrm{hom}}=0$, encode
state-dependent freedom subject to the chosen boundary conditions.

The stress-energy tensor is defined by varying the metric while holding
$\varphi$ fixed, and then imposing \eqref{eq:polyakov-auxiliary}:
\begin{equation}
\begin{split}
&T^{\mathrm P}_{ab}=-\frac{2}{\sqrt{-\gamma}}\frac{\delta S_{\mathrm P}}{\delta g^{ab}}\\
&
 =\frac{\hbar C}{48\pi}
 \left[
 \nabla_a\varphi\nabla_b\varphi
 -\frac12\gamma_{ab}(\nabla\varphi)^2
 -2\nabla_a\nabla_b\varphi+2\gamma_{ab}\Box\varphi
 \right].
 \label{eq:polyakov-stress}
 \end{split}
\end{equation}
It obeys
\begin{equation}
\begin{split}
&\nabla^aT^{\mathrm P}_{ab}
 =\frac{\hbar C}{48\pi}
 (\Box\varphi-R)\nabla_b\varphi=0, \\
&T^{\mathrm P\,a}{}_{a}
 =\frac{\hbar C}{24\pi}R\,.
\end{split}
\end{equation}
Thus its trace reproduces the conformal anomaly.
The dimensions are
\begin{equation}
 [\varphi]=1,\, [\hbar]=ML,\,
 [S_{\mathrm P}]=ML,\, [T^{\mathrm P}_{ab}]=M/L
\end{equation}
in coordinates with dimension $L$.

We study QTG coupled to two-dimensional conformal matter described by the Polyakov action. The total action is
\be
S=S_{\mathrm{QTG}}+S_P\, .
\ee
To obtain dimensionless field equations, we define
\be
\tilde{S}=-\dfrac{G_D}{b\ell^{D-2}}S\hh \tilde{S}_P=-\dfrac{G_D}{b\ell^{D-2}}S_P\, .
\ee
We define the $D$-dimensional Planck length $\ell_{Pl}$ by
\be
\ell_{Pl}=(\hbar G_D)^{1/(D-2)}\, .
\ee
Then the dimensionless total action of our model takes the form
\be
\begin{split}
&\tilde{S}=\tilde{S}_{QTG}+a  \int d^2x\,\sqrt{-\gamma}\,
 \left[(\nabla\varphi)^2+2\varphi R\right]\, ,\\
& a=\dfrac{C}{96\pi b}\big( \dfrac{\ell_{Pl}}{\ell} \big)^{D-2}
=\dfrac{C}{6(D-2)\Omega_{D-2}}\big( \dfrac{\ell_{Pl}}{\ell}\big)^{D-2}
\, .
\end{split}
\ee

The overall minus sign in the definition of $\tilde S$ makes the
Polyakov contribution positive in this expression; it does not change
the field equations. In particular,
\be
 \tilde S_{\mathrm{QTG}}=\int dt\,dr\,N'H.
\ee
All quantities in the following equations are dimensionless. More
explicitly, the rescaling means $r_{\rm phys}=\ell r$,
$t_{\rm phys}=\ell t$, $p_{\rm phys}=\ell^{-2}p$, and
$h_{\rm phys}(p_{\rm phys})=\ell^{-2}h(p)$.
The functions $N$, $f$, and $\varphi$ are dimensionless already.

\subsection{Explicit form of total reduced action}
We take $N>0$ and use a coordinate patch with $f\ne0$.
For the two-dimensional metric above,
\be\n{RR}
\begin{split}
 &\sqrt{-\gamma}=N,\\
 &R=-f''-3\frac{N'}N f'-2f\frac{N''}N
   =-\frac1N\left(Nf'+2fN'\right)'.
\end{split}
\ee
For a general auxiliary field $\varphi(t,r)$,
\be
 (\nabla\varphi)^2=-\frac{\dot\varphi^2}{N^2f} +f(\varphi')^2.
\ee
Consequently, the action evaluated on the static metric is
\be
\begin{split}
 \tilde S=\int dt\,dr\,&\bigg\{N'H
 +a\bigg[-\frac{\dot\varphi^2}{Nf}+Nf(\varphi')^2\\ &-2\varphi\left(Nf'+2fN'\right)'\bigg]\bigg\}.
\end{split}
\ee
Integration by parts in $r$ gives the first-derivative form
\be\n{ACTION}
\begin{split}
 \tilde S=\int dt\,dr\,&\bigg\{N'H
 +a\bigg[-\frac{\dot\varphi^2}{Nf}+Nf(\varphi')^2\\
&+2\varphi'\left(Nf'+2fN'\right)\bigg]\bigg\} ,
\end{split}
\ee
up to the boundary term
$-2a\int dt\,[\varphi(Nf'+2fN')]_{\partial}$.

We now restrict to the static auxiliary-field sector, $\varphi=\varphi(r)$. Define
$q=\varphi'$
then one has
\be\n{RED_LAG}
\begin{split}
&\tilde{S}= \int dt\,dr \, \CAL{L} ,\\
&\CAL{L}=N'H+a\left[Nfq^2+2Nf'q+4fN'q\right].
\end{split}
\ee
The auxiliary field must be varied independently before its equation
of motion is imposed.

\section{Variations and field equations}

\subsection{Equations for $f(r)$, $N(r)$ and $q(r)$.}

Writing the first variation as
\be
\begin{split}
 \delta\tilde S={}&\int dt\,dr\,
 \left(\mathcal E_N\delta N+\mathcal E_f\delta f
                  +\mathcal E_\varphi\delta\varphi\right)\\
 &+\int dt\,[\Theta]_{\partial},
\end{split}
\ee
we obtain
\be\label{eq:EN}
\begin{split}
 &\mathcal E_N=-H'+a\left(fq^2-2f'q-4fq'\right), \\
 &\mathcal E_f=-AN'+a\left(Nq^2+2N'q-2Nq'\right),\\
 &\mathcal E_\varphi=-2a\left(Nfq+Nf'+2fN'\right)'.
\end{split}
\ee
The boundary variation of the first-derivative action is
\be
\begin{split}
 \Theta={}&(H+4afq)\delta N+2aNq\,\delta f\\
 &+2a(Nfq+Nf'+2fN')\delta\varphi.
\end{split}
\ee
It vanishes for fixed endpoint values of $N$, $f$, and $\varphi$.
For other boundary conditions, appropriate boundary terms must be
included. The bulk field equations are
\be
\begin{split}\label{EQ_fNq}
 &H'=a\left(fq^2-2f'q-4fq'\right),\\
 &(A-2aq)N'=aN\left(q^2-2q'\right),\\
&\left(Nfq+Nf'+2fN'\right)'=0.
\end{split}
\ee
The last equation assumes $a\ne0$ and is precisely
\be
 \Box\varphi=\frac1N(Nf\varphi')'=R.
\ee
It has the first integral
\be
 Nfq+Nf'+2fN'=K,
 \label{eq:first-integral}
\ee
where $K$ is a constant. Thus, locally,
\be
 \varphi=-\ln|N^2f|+K\int^r\frac{d\bar r}{N(\bar r)f(\bar r)}
                  +\varphi_0.
\ee
The constants are fixed by boundary and state conditions; they should
not be set to zero without specifying those conditions.

For direct use in terms of $f$ and $h$, the left-hand side of the first equation in \eqref{EQ_fNq} is
\be
 H'=r^{D-2}\left[(D-1)h-2ph_p\right]
                  -r^{D-3}h_p f'.
\ee
Equations \eqref{EQ_fNq}, together with
$p=(1-f)/r^2$ and $q=\varphi'$, form the static system.

As a check, for $a=0$ the gravitational equations reduce to
$H'=0$ and $AN'=0$. On a nondegenerate branch with $h_p\ne0$,
$N$ is constant and can be normalized to unity by rescaling $t$.
The auxiliary field decouples when $a=0$.

The assumption $\varphi=\varphi(r)$ selects a static sector with
$T^{\mathrm P}_{tr}=0$. More general stationary quantum states can
require time-dependent auxiliary fields; the equations above must
then be supplemented or rederived, including the mixed metric
constraint. Finally, the Polyakov term describes two-dimensional
conformal matter. It is not the complete effective action of a
spherically reduced $D$-dimensional scalar field, whose coupling to
the radius introduces additional dilaton-dependent terms.

\subsection{Constraint equation}

Equations \eqref{EQ_fNq} form a first-order system for $f(r)$, $N(r)$ and $q(r)$.
The coefficient matrix for their radial derivatives is singular. Consequently, a linear combination of the three equations eliminates all radial derivatives and yields an algebraic constraint on the field values at each radius. We now derive this constraint.

Later we shall demonstrate that for asymptotically flat solutions one should put the integration constant $K=0$. In what follows we impose this condition. We also assume that $r$ is at a regular radius with $r>0$, $N>0$,
$f\ne0$, $a\ne0$, and $h_p\ne0$. Write
\be
\begin{split}\n{pHAJ}
 &p=\frac{1-f}{r^2},\quad H=r^{D-1}h(p),\\
 &A=r^{D-3}h_p,\quad J=r^{D-2}\big[(D-1)h-2ph_p\big], \\
 &w=N'/N \hh L=A-2aq \, .
\end{split}
\ee
The chain rule gives
\be \n{HHH}
H'=J-Af'\, .
\ee
and the three original \eqref{EQ_fNq} therefore read
\be\n{EQNN}
\begin{split}
& Lf'-4afq' = J-afq^2, \\
& Lw+2aq' = aq^2, \\
& f'+2fw = -fq.
\end{split}
\ee
Add $2f$ times of the second equation to first equation  and subtract $L$ times
third equation. The left-hand side vanishes identically. The same
combination on the right gives
\be \n{CONSTR}
\begin{split}
 C(r,f,q)&\equiv J-afq^2+2afq^2+Lfq\\
   &=J+Afq-afq^2=0\, .
\end{split}
\ee
This constraint indicates that
the derivative matrix has rank two.

In order to explain the degeneracy of a system \eqref{EQNN} and the origin of the constraint equation \eqref{CONSTR} let us write the metric \eqref{METRIC} in a more general form.
Let $x$ be an arbitrary radial parameter and write
\be \n{met_b}
ds^2=-u(x)^2dt^2+b(x)^2dx^2+r(x)^2d\Omega_{D-2}^2\, ,
\ee
on a static patch with $r_x>0$, where a subscript $x$ denotes $d/dx$.
The metric \eqref{METRIC} follows from \eqref{met_b} if one puts
\be \n{fN}
f=(r_x/b)^2\hh N=ub/r_x ,
\ee
In the static auxiliary-field sector the action \eqref{RED_LAG}
becomes
\be \n{EXT_ACT}
\begin{split}
 \widetilde S[u,b,r]=\int dt\,dx\,
 &\left[H N_x+\frac{a}{r_x}
 \left\{Nf\,\varphi_x^2   \right. \right.\\
 &\left. \left.+2\varphi_x(Nf_x+2fN_x)\right\}\right].
\end{split}
\ee
This follows by substituting $dr=r_x\,dx$ and
$\partial_r=r_x^{-1}\partial_x$.
To find a variation of this action with respect to $b$ for fixed $u$ and $r$, one should substitute expressions \eqref{fN} for $f$ and $N$ into \eqref{EXT_ACT}. Note that for this variation one has
\be
\dfrac{\delta N}{N}=\dfrac{\delta b}{b}\hh
\dfrac{\delta f}{f}=-2\dfrac{\delta b}{b}\, .
\ee
Hence, upon choosing $x=r$, its equation is
\be
N\mathcal E_N-2f\mathcal E_f=0\, .
\ee
The terms containing $q'$ cancel,
leaving, with $J=(\partial H/\partial r)_f$,
\[
 (A-2aq)\left(f'+2fN'/N\right)=J+afq^2 .
\]
In the $K=0$ sector the auxiliary-field first integral gives
$f'+2fN'/N=-fq$, so the radial lapse equation reduces to the
algebraic constraint
\be\n{CCC}
\mathcal C=J+Afq-afq^2=0\, .
\ee
Radial reparametrization invariance then makes the $r(x)$ equation
dependent on the other field equations wherever $r_x\ne0$.

\section{First-order system for $H$, $N$, and $q$ with $K=0$}
\label{subsec:HNq}

\subsection{Asymptotic flatness}

In the asymptotically flat region, the metric functions satisfy
\begin{equation}
 f\to1,\qquad N\to N_0=\mathrm{const}>0\,.
\end{equation}
The time coordinate is normalized by introducing $T=N_0t$.
The auxiliary field has the leading asymptotic behavior
\begin{equation}
 \varphi\sim\frac{K}{N_0}r+\varphi_0\,.
\end{equation}
Consequently, in the original coordinates $(t,r)$, the leading term
of the Polyakov stress-energy tensor is
\begin{equation}
 T^{\mathrm P}_{ab}
 \sim\frac{\hbar C K^2}{96\pi}
 \begin{pmatrix}
 1&0\\
 0&N_0^{-2}
 \end{pmatrix}\,.
\end{equation}
Thus, a nonzero $K$ produces a nonvanishing asymptotic energy density
and pressure. To impose an asymptotically empty state, with
$T^{\mathrm P}_{ab}\to0$, we set the integration constant $K=0$.

\subsection{New variables}

We now impose $K=0$, as appropriate to the asymptotically empty
static state discussed below.
For numerical calculations it is  convenient to change the variables $(f,N,q)$ to a new set of variables $(H,N,q)$. We explain now how
\eqref{EQNN} give three first-order
ordinary differential equations for $H(r)$, $N(r)$, and $q(r)$,
where $H(r)$ is defined in \eqref{pHAJ}.
Let us note that $f$ is reconstructed algebraically from $r$ and $H$, rather
than evolved as a fourth independent variable. On a chosen locally
invertible branch of $h$,
\begin{equation}
 p=h^{-1}\!\left(\frac{H}{r^{D-1}}\right),
 \qquad f=1-r^2p.
 \label{eq:HNq-inverse}
\end{equation}
The symbol $h^{-1}$ denotes the inverse function, not $1/h$.
All quantities are dimensionless, and a prime still denotes $d/dr$.
We assume $a\ne0$, $r>0$, $N>0$, $f\ne0$, and $h_p\ne0$.
The center and horizons require limits or a separate regular
coordinate formulation.

Taking partial derivatives of the constraint function $\CAL{C}$ defined in \eqref{CCC} with respect to the indicated argument gives
\be
\begin{split}
 \mathcal C_r&=J_r+fqA_r,\\
 \mathcal C_f&=J_f+q(A+fA_f)-aq^2,\\
 \mathcal C_q&=fL.
\end{split}
\ee
For explicit evaluation, define
\begin{equation}
\begin{split}
 &U=(D-1)h-2ph_p,\\
 &V=\frac{dU}{dp}=(D-3)h_p-2ph_{pp}.
\end{split}
\end{equation}
Then
\be
\begin{split}
 A_r&=r^{D-4}V,\\
 A_f&=-r^{D-5}h_{pp},\\
 J_r&=r^{D-3}\big[(D-2)U-2pV\big],\\
 J_f&=-r^{D-4}V.
\end{split}
\ee

\subsection{Determining the lapse derivative}
With $w=N'/N$, system \eqref{EQNN} gives
\begin{align}
 f'&=-f(q+2w), \label{eq:HNq-fprime}\\
 q'&=\frac{q^2}{2}-\frac{L}{2a}w. \label{eq:HNq-qprime}
\end{align}
On substituting these expressions into the first equation in \eqref{EQ_fNq}, we find
\begin{equation}
 H'=afq^2+2Afw. \label{eq:HNq-Hprime2}
\end{equation}
Equating this to $J-Af'=J+Afq+2Afw$ recovers
\begin{equation}
 \mathcal C=0. \label{eq:HNq-Czero}
\end{equation}
Differentiation along a solution now determines $w$:
\begin{equation}
\begin{split}
 0&=\mathcal C_r+\mathcal C_f f'+fLq'\\
  &=\mathcal C_r-fq\mathcal C_f+\tfrac12 fLq^2
     -\frac{f}{2a}(L^2+4a\mathcal C_f)w .
\end{split}
\end{equation}
Thus, wherever $f\Delta\ne0$,
\be
\begin{split}\n{WW}
 \Delta&=L^2+4a\mathcal C_f,\\
 w=W(r,H,q)&=\frac{a\big(2\mathcal C_r-2fq\mathcal C_f+fLq^2\big)}      {f\Delta}.
\end{split}
\ee
Here $f=f(r,H)$ is given by \eqref{eq:HNq-inverse}. Substitution of
$w=W$ into $H'=J-Af'$, $N'=Nw$, and \eqref{eq:HNq-qprime} yields
\eqref{eq:HNq-system}.

\subsection{The closed first-order system.}
The desired equations can now be written as
\begin{equation}
\begin{split}
 H'&=J+Af\big(q+2W\big),\\
 N'&=NW,\\
 q'&=\frac{q^2}{2}-\frac{A-2aq}{2a}W.
 \end{split}
 \label{eq:HNq-system}
\end{equation}
Equations \eqref{pHAJ} and \eqref{WW},
specify all coefficients.
On the constraint surface, the first equation is equivalently
$H'=afq^2+2AfW$. The displayed form ensures that the reconstructed
$f$ satisfies \eqref{eq:HNq-fprime} directly.

At an initial radius $r_*>0$, choose $H_*$ on the selected inverse
branch, reconstruct $f_*$, and choose $q_*$ so that
$\mathcal C(r_*,f_*,q_*)=0$. For example, the possible real roots are
\begin{equation}
 q_* = \frac{A_*\pm\sqrt{A_*^2+4aJ_*/f_*}}{2a}.
 \label{eq:HNq-initialroots}
\end{equation}
A root must be chosen consistently with the desired solution and
boundary conditions. The value $N_*>0$ fixes the normalization of
time. These are three first-order equations with constrained initial
data, not three freely specifiable initial values. By construction,
$\mathcal C'=0$ under \eqref{eq:HNq-system}; thus imposing
$\mathcal C=0$ initially preserves the original metric equation.
It is also a useful residual to monitor in numerical integration.

The explicit system requires $\Delta\ne0$. If $\Delta=0$, the
constraint derivative no longer determines $N'/N$ in this form,
and one must return to the undivided equations and examine
compatibility. This is not, by itself, proof of a curvature
singularity. In particular, no division by $A-2aq$ was used, so
$A-2aq=0$ alone does not invalidate the system when $\Delta\ne0$.
The limit $a=0$ must be treated separately as described above.

Finally, the auxiliary field itself is recovered by quadrature,
\begin{equation}
\varphi(r)=\varphi_*+\int_{r_*}^{r}q(s)\,ds =-\ln|N^2f|+\varphi_0.
\end{equation}
For $K=0$, the equations for $H$ and $q$ do not depend on the
normalization of $N$; after solving them,
$N=N_*\exp\big(\int_{r_*}^{r}W(s)\,ds\big)$.

\section{Hayward--type QTG model}
\label{sec:hayward}

\subsection{Vacuum solutions for $a=0$}

We first consider spherically symmetric vacuum solutions in $D\ge 4$
dimensions. For $a=0$, one can set $N=1$ by a constant rescaling
of the time coordinate. For the Hayward-type model,
\begin{equation}
 h(p)=\frac{p}{1-p}, \,
 h_p=\frac{1}{(1-p)^2},\,
 h_{pp}=\frac{2}{(1-p)^3}.
\end{equation}
Integrating the equation $H'=0$ gives
\be
H=\mu\hh h=\dfrac{\mu}{r^{D-1}}\, .
\ee
The corresponding dimensionful integration constant
$\mu_{\mathrm{phys}}$ is related to the mass $M$ measured at infinity by
\be
\mu_{\mathrm{phys}}
=\dfrac{16\pi G_D M}{(D-2)\Omega_{D-2}}\, ,
\ee
where $\Omega_{D-2}$ is the area of the unit $(D-2)$-sphere.
It has dimensions $[\mu_{\mathrm{phys}}]=L^{D-3}$.
In $D=4$, $\mu_{\mathrm{phys}}=2GM$.
In the units adopted in this paper, we use the dimensionless
quantities $\mu=\mu_{\mathrm{phys}}/\ell^{D-3}$ and
$r=r_{\mathrm{phys}}/\ell$.

The metric function $f$ is determined by
\be \n{pppp}
f=1-r^2p\hh
p=\dfrac{h}{1+h}
=\dfrac{\mu}{\mu+r^{D-1}}\, .
\ee
For $\mu>0$, it has the universal de Sitter behavior near the center,
\be
f=1-r^2+\dfrac{r^{D+1}}{\mu}
+O(r^{2D})\, .
\ee
In terms of the physical radial coordinate, the leading terms are
$f=1-r_{\mathrm{phys}}^2/\ell^2+\ldots$.

There exists a critical value $\mu=\mu_*$ that separates solutions
with two horizons ($\mu>\mu_*$) from horizonless solutions
($0<\mu<\mu_*$). At $\mu=\mu_*$, the two horizons merge into
a degenerate horizon at $r=r_*$. The critical parameters satisfy
\be
f(r_*)=0\hh f'(r_*)=0\, .
\ee
Denoting the values of the primary curvature invariant and its radial
derivative at $r=r_*$ by $p_*$ and $p'_0$, respectively, we obtain
\be
p_*=\dfrac{1}{r_*^2}\hh
p'_*=-\dfrac{2}{r_*^3}\, .
\ee
Using \eqref{pppp}, these conditions yield
\be
\begin{split}
&(1-r_*^2)\mu_*=-r_*^{D-1}\, ,\\
&\mu_*=\dfrac{D-1}{2}r_*^{D-3}\, .
\end{split}
\ee
Solving these equations gives
\be
r_*=\sqrt{\dfrac{D-1}{D-3}}\hh
\mu_*=
\dfrac{(D-1)^{(D-1)/2}}
{2(D-3)^{(D-3)/2}}\, .
\ee
For $D=4$, these expressions reduce to
\be
r_*=\sqrt{3}\hh
\mu_*=\dfrac{3\sqrt{3}}{2}\, .
\ee

\subsection{Algebraic reconstruction and first-order equations}

Throughout this section $K=0$, $D\ge4$, and all quantities are
dimensionless. A prime denotes $d/dr$.

Introduce the algebraic quantities
\begin{equation}
 x=\frac{H}{r^{D-1}},\qquad z=1+x.
 \label{eq:hay-xz}
\end{equation}
The curvature variable and metric function are then
\begin{equation}
\begin{split}
 p&=\frac{H}{r^{D-1}+H}=\frac{x}{z},\\
 f&=1-\frac{r^2H}{r^{D-1}+H}
   =1-r^2+\frac{r^{D+1}}{H+r^{D-1}}.
\end{split}
\label{eq:hay-reconstruct}
\end{equation}
Thus neither $p$ nor $f$ is an additional unknown. The branch
connected to small curvature has $z>0$, or equivalently $p<1$.

Only two coefficients from Sec.~3 are needed:
\begin{equation}
 A=r^{D-3}z^2,\qquad J=\frac{H}{r}(D-3-2x).
 \label{eq:hay-AJL}
\end{equation}
Define
\begin{equation}
 P_D(x)=(D-3)(D-4)+(18-4D)x+8x^2,
\end{equation}
and the following explicit functions of $r,H,q$:
\begin{equation}
\begin{split}
 E={}&A-\frac{4a(D-3-4x)}r-\frac{8afzq}{r^2},\\
 Q={}&\frac{2xP_D(x)}{fz^2}+\frac{4(D-3-4x)q}{r}\\
 &+\left(\frac{4fz}{r^2}-1\right)q^2.
\end{split}
\label{eq:hay-EQ}
\end{equation}
Here $Q$ is an algebraic coefficient, not an electric charge.
Substitution in the general formulas of Sec.~3 gives
$\Delta=AE$ and $W=aQ/E$. In particular, all cubic powers of $q$
in the numerator of $W$ cancel.

On the constraint surface, the evolution equations have the same
compact form for every $D$:
\begin{equation}
\begin{aligned}
H'&=af\left(q^2+\frac{2AQ}{E}\right),\\
 N'&=\frac{aQ}{E}N,\\
 q'&=\frac{q^2}{2}-\frac{(A-2aq)Q}{2E}.
 \end{aligned}
 \label{eq:hay-system}
\end{equation}
They must be supplemented by the algebraic constraint \eqref{CCC}.
This is the specialized form of the constraint in Sec.~3, rather
than an independent evolution equation. The reconstructed metric
satisfies $f'=-f(q+2aQ/E)$ on this surface. If an evolution form
with $\mathcal C'=0$ also away from the constraint surface is
desired, replace the first equation by
$H'=J+Af(q+2aQ/E)$, as in Sec.~3.

For $D=4$, simply set
\begin{equation}
\begin{split}
 x&=\frac{H}{r^3},\qquad A=rz^2,\\
 J&=\frac{H}{r}(1-2x),\qquad P_4=2x(1+4x).
\end{split}
\end{equation}
No separate four-dimensional evolution system is required.

\subsection{Initial data and domain of validity}
At a positive initial radius, the constraint fixes $q$ to one of
its real roots,
\begin{equation}
 q_\pm=\frac{A}{2a}\left(1\pm\sqrt{1+\frac{4aJ}{fA^2}}\right).
\end{equation}
Since $A>0$, the root with a finite $a\to0$ limit is conveniently
written as
\begin{equation}
 q_-=-\frac{2J}{fA\left(1+\sqrt{1+4aJ/(fA^2)}\right)}
       \longrightarrow-\frac{J}{Af}.
\end{equation}
The equations assume $a\ne0$, $r>0$, $N>0$, $z\ne0$, $f\ne0$,
and $E\ne0$. Because $\Delta=AE$, the degeneracy condition is
$E=0$; $A-2aq=0$ alone is not an obstruction. The center and
horizons must be treated by limits or the undivided equations.

Once $H$ and $q$ are known, the remaining functions follow from
\begin{align}
 N(r)&=N_*\exp\left[\int_{r_*}^{r}\frac{aQ(s)}{E(s)}\,ds\right],\\
 \varphi(r)&=\varphi_*+\int_{r_*}^{r}q(s)\,ds
           =-\ln|N^2f|+\varphi_0.
\end{align}
The normalization $N_*$ does not affect $H$ or $q$.
For $a=0$, the auxiliary field decouples, and the gravitational
solution is
\begin{equation}
 H=\mu,\qquad N=N_0,\qquad
 f=1-\frac{\mu r^2}{r^{D-1}+\mu}.
\end{equation}
For $D=4$, identifying $\mu=2M$ gives the usual Hayward metric.

\subsection{Central expansion}
\label{subsec:hay-center}
Consider the limiting data
\begin{equation}
 N(0)=1,\qquad H(0)=H_0\ne0,\qquad q(0)=0.
 \label{eq:hay-center-data}
\end{equation}
The right-hand sides of the explicit system are singular at $r=0$,
so these conditions are imposed through a series as $r\to0^+$.
A simple derivation uses the undivided equations
\begin{align}
 H'&=a(fq^2-2f'q-4fq'),\label{eq:center-recH}\\
 (A-2aq)w&=a(q^2-2q'),\qquad w=N'/N,
 \label{eq:center-recw}\\
 q&=-f'/f-2w.\label{eq:center-recq}
\end{align}
For this model,
\begin{equation}
 A=\frac{(H+r^{D-1})^2}{r^{D+1}}
   \sim\frac{H_0^2}{r^{D+1}}.
\end{equation}
Together with \eqref{eq:hay-reconstruct}, this gives
\begin{equation}
\begin{split}
 q&=2r+O(r^3),\qquad H'=-8a+O(r^2),\\
 w&=-\frac{4a}{H_0^2}r^{D+1}+O(r^{D+2}).
\end{split}
\end{equation}
The leading terms in the constraint cancel as
$J\sim-2H_0^2r^{-D}$ and $Afq\sim2H_0^2r^{-D}$.
The other root, $q\sim H_0^2/(ar^{D+1})$, is excluded by $q(0)=0$.

The dimension-independent de Sitter contributions can be kept in
closed form, making the series especially compact:
\begin{equation}
\begin{split}
 &H(r)=H_0-4a\big(r+\arctanh r\big)\\
 &+\frac{4a(D+1)}{H_0}r^D
 +\frac{32a^2(D+1)}{H_0^2}r^{D+1}
 +O(r^{D+2}),\\
 &q(r)=\frac{2r}{1-r^2}-\frac{D+1}{H_0}r^D\\
 &-\frac{8a(D+1)}{H_0^2}r^{D+1}+O(r^{D+2}),\\
 &N(r)={}1-\frac{4a}{(D+2)H_0^2}r^{D+2}\\
 &-\frac{64a^2}{(D+3)H_0^3}r^{D+3}+O(r^{D+4}).
\end{split}
\label{eq:center-N-general}
\end{equation}
These are local expansions for fixed $H_0\ne0$ and $D\ge4$.
The functions $\arctanh r=r+r^3/3+r^5/5+\cdots$ and
$2r/(1-r^2)=2r+2r^3+2r^5+\cdots$ merely collect the universal
Taylor terms. They do not establish the behavior of the full
solution at $r=1$ or beyond the local domain of the expansion.

The metric and curvature variable are
\begin{align}
 f(r)&=1-r^2+\frac{r^{D+1}}{H_0}
             +\frac{8a}{H_0^2}r^{D+2}+O(r^{D+3}),\\
 p(r)&=1-\frac{r^{D-1}}{H_0}
             -\frac{8a}{H_0^2}r^D+O(r^{D+1}).
 \label{eq:center-fp}
\end{align}
Thus the core is de Sitter-like, with $p(0)=1$ and two-dimensional
scalar curvature $R(0)=2$. The pole of $h(p)$ is compensated by
$r^{D-1}$ in $H=r^{D-1}h(p)$. The small-curvature branch requires
$H_0>0$; $H_0<0$ approaches the pole from $p>1$.
Finite limiting curvature does not imply arbitrary-order
smoothness in Cartesian coordinates.

The auxiliary field is
\begin{equation}
\begin{split}
 \varphi(r)={}&\varphi_c-\ln(1-r^2)-\frac{r^{D+1}}{H_0}\\
 &-\frac{8a(D+1)}{(D+2)H_0^2}r^{D+2}+O(r^{D+3}).
\end{split}
\end{equation}
These series provide starting values at a small positive radius
for numerical integration. They are not uniform as $H_0\to0$;
that case requires a different central expansion.

Asymptotic flatness and the asymptotically empty $K=0$ state also
give $fq\to0$ at infinity. Consequently,
\begin{equation}
 m_\infty=\frac{H(0)}2
 +\frac a2\int_0^\infty\left(fq^2+2f'q\right)\,dr .
 \label{eq:mass-integral}
\end{equation}

Equation~\eqref{eq:mass-integral} shows that the asymptotic mass
combines the central parameter with an integrated matter contribution.
The central critical parameter $H_{0,\mathrm{crit}}(a)$ is itself
nonlinear and nonmonotonic. The numerical near-linearity of
$m_\infty(a)$ therefore corresponds to a substantial cancellation
between the nonlinear dependence of the two terms in
\eqref{eq:mass-integral}.

For comparison, expanding the universal terms above at $D=4$
reproduces
\begin{equation}
\begin{split}
 &H(r)=H_0-8ar-\frac{4a}{3}r^3+\frac{20a}{H_0}r^4\\
 &\quad +\left(-\frac{4a}{5}+\frac{160a^2}{H_0^2}\right)r^5
 +O(r^6), \\
 &q(r)=2r+2r^3-\frac{5}{H_0}r^4
 +\left(2-\frac{40a}{H_0^2}\right)r^5+O(r^6),\\
 &N(r)=1-\frac{2a}{3H_0^2}r^6
          -\frac{64a^2}{7H_0^3}r^7+O(r^8).
\end{split}
 \label{eq:center-N}
\end{equation}
In all dimensions $H'(0)=-8a$, whereas the first correction to the
lapse occurs at order $r^{D+2}$. A different positive value of
$N(0)$ simply multiplies the lapse series by that constant.

\subsection{Near-critical minimons}

We continue to consider the $D=4$ case. In what follows, we denote
$N_0=N(0)$, $N_{\infty}=\lim_{r\to\infty}N(r)$, and
$\nu=N_{\infty}/N_0$. The freedom to rescale the time coordinate
allows a common rescaling of $N_0$ and $N_{\infty}$, while their
ratio $\nu$ remains invariant.

The parameter $]nu$ has a simple physical meaning. Suppose one send a photon of frequency $\omega_{\infty}$. Then when this photon reaches the center of the minimon it frequency will be $\omega_0=\nu \omega_{\infty}$.

For numerical integration, we may
therefore set $N_0=1$. With this normalization, the system of equations
\eqref{eq:hay-system} requires only one initial-data parameter.
A convenient choice is the central value of $H(r)$, which we denote
by $H_0=H(0)$.

\begin{figure}[!htb]%
    \centering \includegraphics[width=0.4\textwidth]{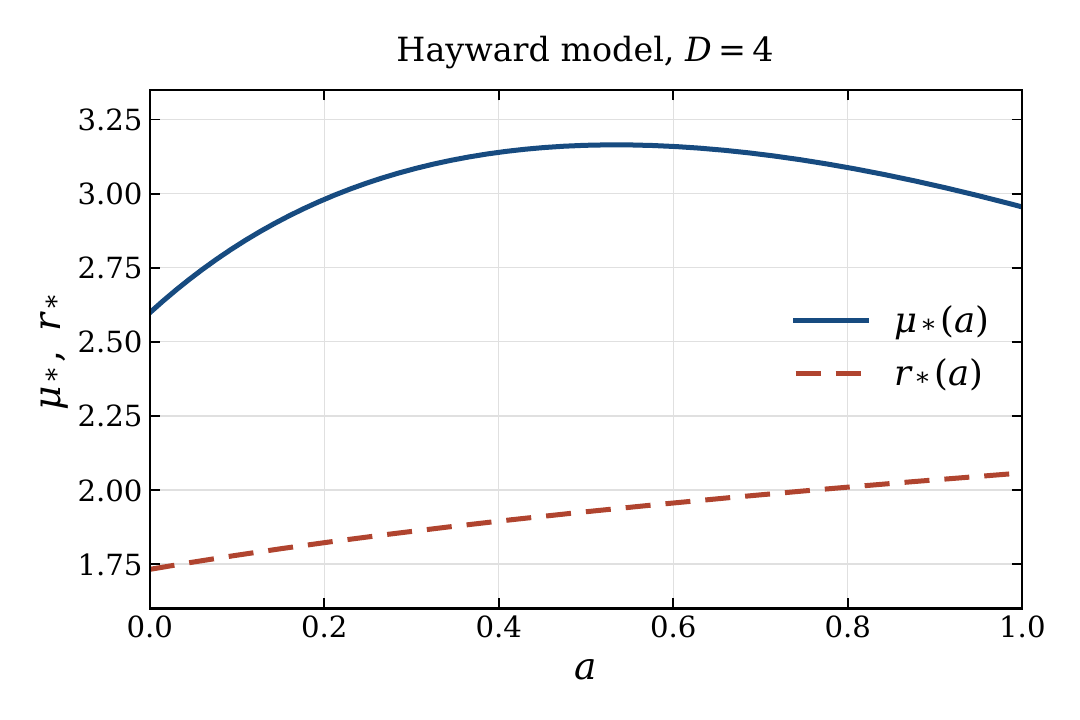}\\[0pt]
\caption{$\mu_*$ and $r_*$ as a function of $a$ for Hayward-type QTG model.}
    \label{F1}
\end{figure}

For $a=0$, there is a critical value $\mu_*=3\sqrt{3}/2$,
such that minimon solutions exist only for $H_0<\mu_*$.
This property persists for $a>0$, but the critical value now depends
on $a$: $\mu_*=\mu_*(a)$. This dependence is
illustrated in figure~\ref{F1}.

The quantity $m_{\infty}=H_{\infty}/2$ determines the asymptotic
mass of a minimon as measured at infinity. Figure~\ref{F2}
shows $m_{\infty}$ as a function of $a$ for extremal minimons.
The mass decreases approximately linearly with $a$.

\vspace{0.5cm}

{ }

\begin{figure}[!htb]%
    \centering \includegraphics[width=0.35\textwidth]{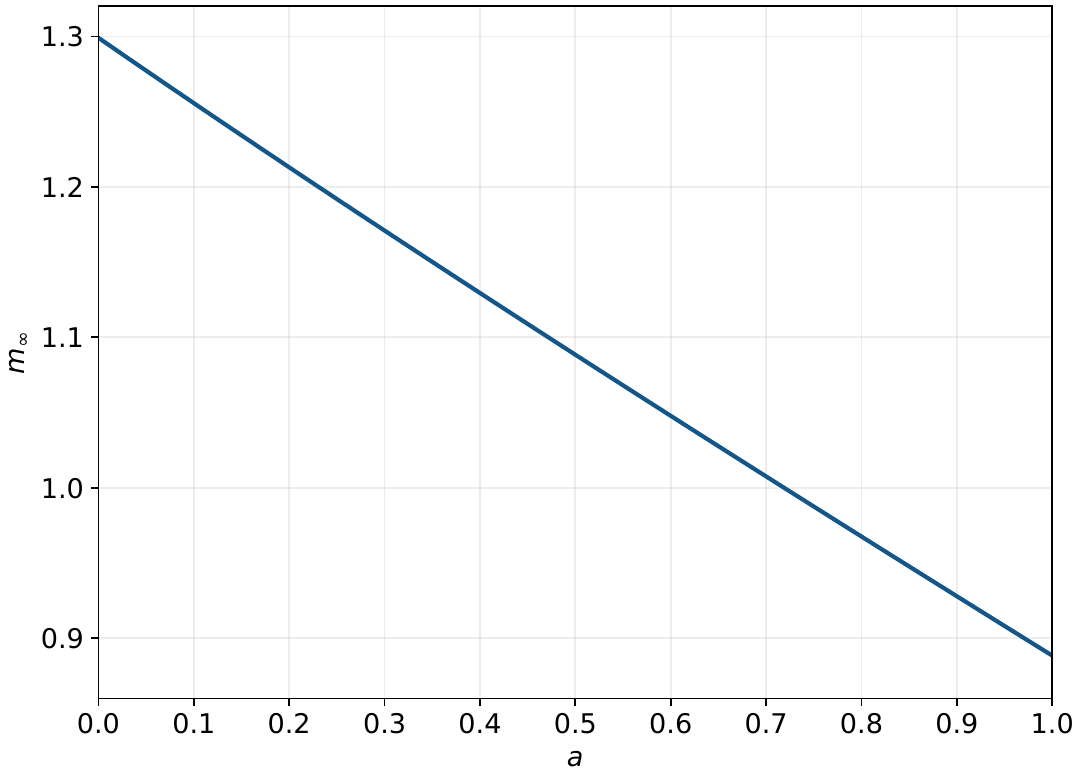}\\[0pt]
\caption{Mass at infinity $m_{\infty}$ of an extremal minimons of the Hayward-type QTG model as a function of $a$.}
    \label{F2}
\end{figure}

Both $\mu_*$ and $m_{\infty}$ are well defined for near-critical
minimons and remain finite in the extremal limit.

\begin{figure}[!htb]%
    \centering \includegraphics[width=0.4\textwidth]{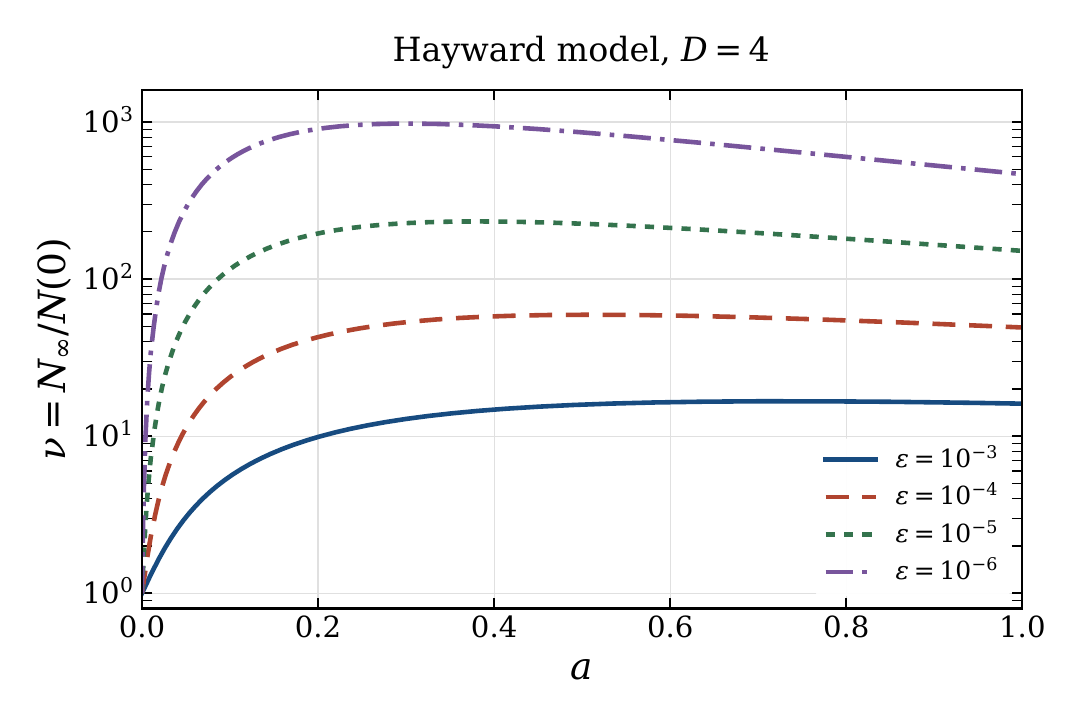}\\[0pt]
\caption{$\nu=N_{\infty}/N_0$ as a function of $a$ for different choice of a small parameter $\epsilon$}
    \label{F3}
\end{figure}

The calculations show that this is not the case for the parameter
$\nu$. Let $r=r_*$ denote the radius at which the metric function
$f(r)$ attains its minimum, $f_*=f(r_*)$, and suppose that
$f_*=\epsilon\ll1$. The numerical results shown in
figure~\ref{F3} indicate that $\nu$ grows without bound as
$\epsilon\to0$ for fixed $a>0$ in the range considered.
The case $a=0$ is exceptional: $N(r)$ is constant and $\nu=1$.

This property implies that a photon arriving at the center from infinity can experience a strong blueshift even for relatively small values of the parameter \(a\).

\vspace{0.5cm}

{ }

\begin{figure}[!htb]%
    \centering \includegraphics[width=0.4\textwidth]{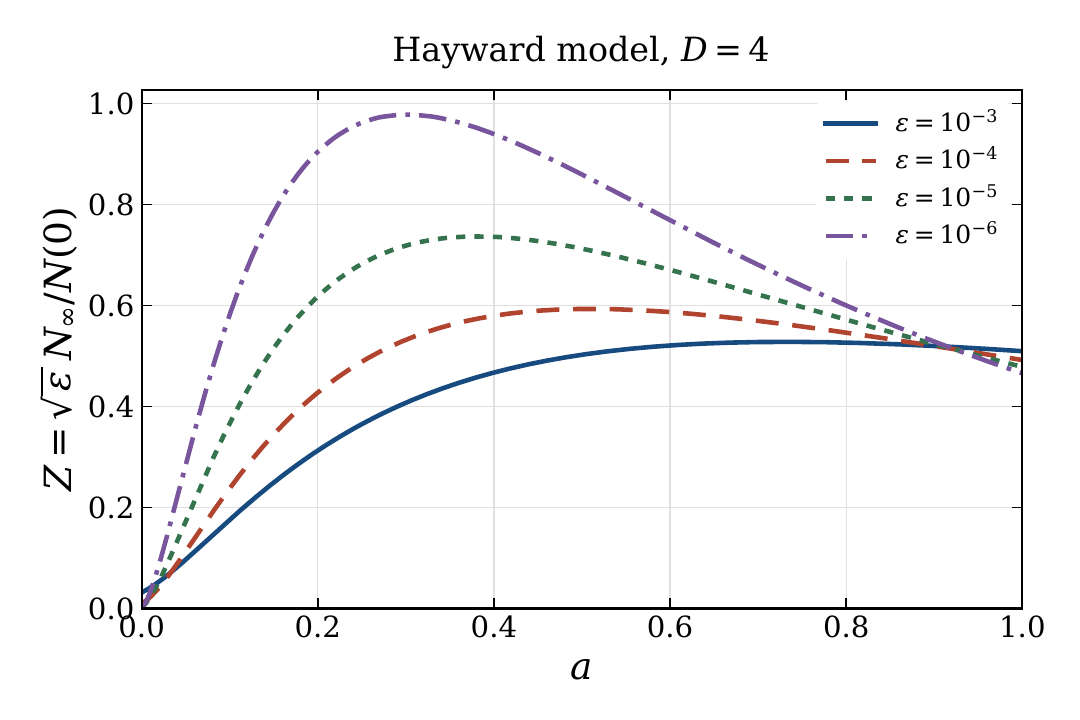}\\[0pt]
\caption{$Z=\sqrt{\epsilon}N_{\infty}/N_0$ as a function of $a$ for different choice of a small parameter $\epsilon$}
    \label{F4}
\end{figure}

The rapid increase in $N(r)$ occurs near the minimum of $f(r)$.
A useful estimate of the enhancement across this region is
$1/\sqrt{\epsilon}$. To examine this estimate, we introduce
$Z=\sqrt{\epsilon}\,\nu$. Figure~\ref{F4} shows $Z$ as a
function of $a$ for several small values of $\epsilon$.
For the values of $\epsilon$ plotted, $Z$ remains of order unity
over much of the displayed range of $a$. However, it retains a
dependence on $\epsilon$, so the figure does not establish an exact
$\nu\propto\epsilon^{-1/2}$ asymptotic scaling.

{ }

\section{Born-Infeld--type QTG model}

\subsection{Vacuum solutions for $a=0$}

We first consider spherically symmetric vacuum solutions in $D\ge 4$
dimensions. For $a=0$, one can set $N=1$ by a constant rescaling
of the time coordinate. In units in which the characteristic length
is unity, the Born--Infeld-type model is defined by
\begin{equation}
\begin{split}
&h(p)=\frac{p}{\sqrt{1-p^2}}, \qquad
 h_p=\frac{1}{(1-p^2)^{3/2}}, \\
 &h_{pp}=\frac{3p}{(1-p^2)^{5/2}}.
\end{split}
\end{equation}
Integrating the equation $H'=0$ gives
\be
H=\mu\hh h=\dfrac{\mu}{r^{D-1}}\, .
\ee
As before, $\mu$ is an integration constant proportional to the mass
$M$ measured at infinity.

The metric function $f$ is determined by
\be \n{pppp}
f=1-r^2p\hh
p=\dfrac{\mu}{\sqrt{\mu^2+r^{2(D-1)}}}\, .
\ee
For $\mu>0$, it has the universal de Sitter behavior near the center,
\be
f=1-r^2+\dfrac{r^{2D}}{2\mu^2}
  +O\bigl(r^{4D-2}\bigr)\, .
\ee

There exists a critical value $\mu=\mu_*$ that separates solutions
with two horizons ($\mu>\mu_*$) from horizonless solutions
($0<\mu<\mu_*$). At $\mu=\mu_*$, the two horizons merge into
a degenerate horizon at $r=r_*$. The critical parameters satisfy
\be
f(r_*)=0\hh f'(r_*)=0\, .
\ee
Denoting the values of the primary curvature invariant and its radial
derivative at $r=r_*$ by $p_*$ and $p'_*$, respectively, we obtain
\be
p_*=\dfrac{1}{r_*^2}\hh
p'_*=-\dfrac{2}{r_*^3}\, .
\ee
Using \eqref{pppp}, these conditions yield
\be
r_*=\left(\dfrac{D-1}{D-3}\right)^{1/4}\hh
\mu_*=\dfrac{1}{\sqrt{2}}\,
\dfrac{(D-1)^{(D-1)/4}}{(D-3)^{(D-3)/4}}\, .
\ee
For $D=4$, these expressions reduce to
\be
r_*=3^{1/4}\hh
\mu_*=\dfrac{3^{3/4}}{\sqrt{2}}\, .
\ee

\subsection{Algebraic reconstruction and first-order equations}

We now take $a>0$, $K=0$, and $D\ge4$. Define
\begin{equation}
 x=\frac{H}{r^{D-1}},\qquad z=\sqrt{1+x^2}.
 \label{eq:bi-xz}
\end{equation}
On the real Born--Infeld branch,
\begin{equation}
 p=\frac{x}{z}=\frac{H}{\sqrt{H^2+r^{2(D-1)}}},\qquad
 f=1-r^2\frac{x}{z}.
 \label{eq:bi-reconstruct}
\end{equation}
Thus $|p|<1$ at every positive radius where $H$ is finite.
Unlike the Hayward model, this inverse is real and single-valued
for either sign of $H$.

The coefficients entering the general first-order system are
\begin{equation}
 A=r^{D-3}z^3,\qquad J=\frac{H}{r}(D-3-2x^2).
 \label{eq:bi-AJ}
\end{equation}
For compactness, put
\begin{equation}
\begin{split}
 B_D&=D-3-6x^2,\\
 P_D&=(D-3)(D-4)+(22-4D)x^2+12x^4,
\end{split}
\end{equation}
and define
\begin{equation}
\begin{split}
 E={}&A-\frac{4aB_D}{r}-\frac{12afxzq}{r^2},\\
 Q={}&\frac{2xP_D}{fz^3}+\frac{4B_Dq}{r}
       +\left(\frac{6fxz}{r^2}-1\right)q^2.
\end{split}
\label{eq:bi-EQ}
\end{equation}
Here $Q$ is an algebraic coefficient. Substitution in the general formulas of Sec.~3 gives $\Delta=AE$ and $W=aQ/E$.
On the constraint surface the equations become
\begin{equation}
\begin{aligned}
 H'&=af\left(q^2+\frac{2AQ}{E}\right),\\
 N'&=\frac{aQ}{E}N,\\
 q'&=\frac{q^2}{2}-\frac{(A-2aq)Q}{2E},
\end{aligned}
\label{eq:bi-system}
\end{equation}
with
\begin{equation}
 \mathcal C=J+Afq-afq^2=0.
 \label{eq:bi-constraint}
\end{equation}
The reconstructed metric obeys $f'=-f(q+2aQ/E)$.
As in the Hayward case, the alternative equation
$H'=J+Af(q+2aQ/E)$ preserves $\mathcal C'=0$ also off the
constraint surface.
For $D=4$, one simply uses $A=rz^3$, $B_4=1-6x^2$, and
$P_4=6x^2(1+2x^2)$.

\subsection{Initial data and central expansion}

At a positive initial radius, the root connected continuously to
$a=0$ is
\begin{equation}
 q_-=-\frac{2J}{fA\left(1+\sqrt{1+4aJ/(fA^2)}\right)}.
 \label{eq:bi-root}
\end{equation}
The square root must be real. The explicit system requires
$r>0$, $N>0$, $f\ne0$, and $E\ne0$.
The condition $A-2aq=0$ alone is not a degeneracy of that system.
The center and any limiting horizon require a separate limiting
analysis.

For the positive-mass central branch, impose
\begin{equation}
 H(0)=H_0>0,\qquad N(0)=1,\qquad q(0)=0.
\end{equation}
Then $A\sim H_0^3r^{-2D}$, and the undivided equations
\eqref{eq:center-recH}--\eqref{eq:center-recq} give
\begin{equation}
\begin{split}
 H(r)={}&H_0-4a\big(r+\arctanh r\big)\\
 &+\frac{4aD}{H_0^2}r^{2D-1}
  +\frac{64a^2D}{H_0^3}r^{2D}+O(r^{2D+1}),\\
 q(r)={}&\frac{2r}{1-r^2}-\frac{D}{H_0^2}r^{2D-1}
  -\frac{16aD}{H_0^3}r^{2D}+O(r^{2D+1}),\\
 N(r)={}&1-\frac{4a}{(2D+1)H_0^3}r^{2D+1}\\
 &-\frac{48a^2}{(D+1)H_0^4}r^{2D+2}+O(r^{2D+3}).
\end{split}
\label{eq:bi-center}
\end{equation}
As before, the rational and inverse hyperbolic functions only
collect the universal local Taylor terms; these expressions do
not describe the full solution near $r=1$.
In particular, $H'(0)=-8a$ in every dimension.

The corresponding metric, curvature variable, and auxiliary field are
\begin{equation}
\begin{split}
 f(r)={}&1-r^2+\frac{r^{2D}}{2H_0^2}
                  +\frac{8a}{H_0^3}r^{2D+1}+O(r^{2D+2}),\\
 p(r)={}&1-\frac{r^{2D-2}}{2H_0^2}
                  -\frac{8a}{H_0^3}r^{2D-1}+O(r^{2D}),\\
 \varphi(r)={}&\varphi_c-\ln(1-r^2)-\frac{r^{2D}}{2H_0^2}\\
 &-\frac{16aD}{(2D+1)H_0^3}r^{2D+1}+O(r^{2D+2}).
\end{split}
\label{eq:bi-center-metric}
\end{equation}
Thus $p(0)=1$ and $R(0)=2$: the core is de Sitter-like.
The first model-dependent correction to $f$ occurs at order
$r^{2D}$, compared with $r^{D+1}$ for Hayward.
The divergence of $h$ at $p=1$ is compensated by $r^{D-1}$ in
$H$. These expansions apply for fixed $H_0>0$ and are not uniform
as $H_0\to0$; they imply finite central curvature, not
arbitrary-order Cartesian smoothness.

For $D=4$, the central series take the concise form
\begin{equation}
\begin{split}
 H={}&H_0-4a(r+\arctanh r)
       +\frac{16a}{H_0^2}r^7+\frac{256a^2}{H_0^3}r^8+O(r^9),\\
 q={}&\frac{2r}{1-r^2}-\frac{4}{H_0^2}r^7
                   -\frac{64a}{H_0^3}r^8+O(r^9),\\
 N={}&1-\frac{4a}{9H_0^3}r^9
        -\frac{48a^2}{5H_0^4}r^{10}+O(r^{11}).
\end{split}
\label{eq:bi-center4}
\end{equation}
They provide initial values at a small positive radius.
An arbitrary positive $N(0)$ multiplies the lapse by that constant.

\subsection{Asymptotic mass and near-critical solutions}

We restrict the following discussion to $D=4$.
For a solution reaching an asymptotically flat end, write
$H_\infty=2m_\infty$ and $N_\infty=\lim_{r\to\infty}N$.
The leading asymptotic terms are
\begin{equation}
\begin{split}
 H&=H_\infty+\frac{4aH_\infty}{r^2}+O(r^{-3}),\\
 f&=1-\frac{H_\infty}{r}-\frac{4aH_\infty}{r^3}+O(r^{-4}),\\
 N&=N_\infty\left[1+\frac{4aH_\infty}{3r^3}+O(r^{-4})\right],\\
 q&=-\frac{H_\infty}{r^2}+O(r^{-3}).
\end{split}
\label{eq:bi-infinity}
\end{equation}
Integrating the $H$ equation gives the same mass identity as for
Hayward,
\begin{equation}
 m_\infty=\frac{H_0}{2}
 +\frac a2\int_0^\infty(fq^2+2f'q)\,\dd r,
\end{equation}
where the endpoint term $fq$ vanishes at both ends.
Moreover,
\begin{equation}
 \ln\nu=\int_0^\infty\frac{aQ}{E}\,\dd r,
 \qquad \nu=\frac{N_\infty}{N(0)}.
\end{equation}
The central photon frequency is $\omega_0=\nu\omega_\infty$.
At $a=0$, $\mu_*=3^{3/4}/\sqrt2$,
$m_{\infty,\mathrm{crit}}=3^{3/4}/(2\sqrt2)$, and $\nu=1$.
For $a>0$, the central critical parameter and asymptotic mass
must be obtained from the global solution, rather than from the
vacuum horizon formula.

\begin{figure}[!htb]%
    \centering \includegraphics[width=0.4\textwidth]{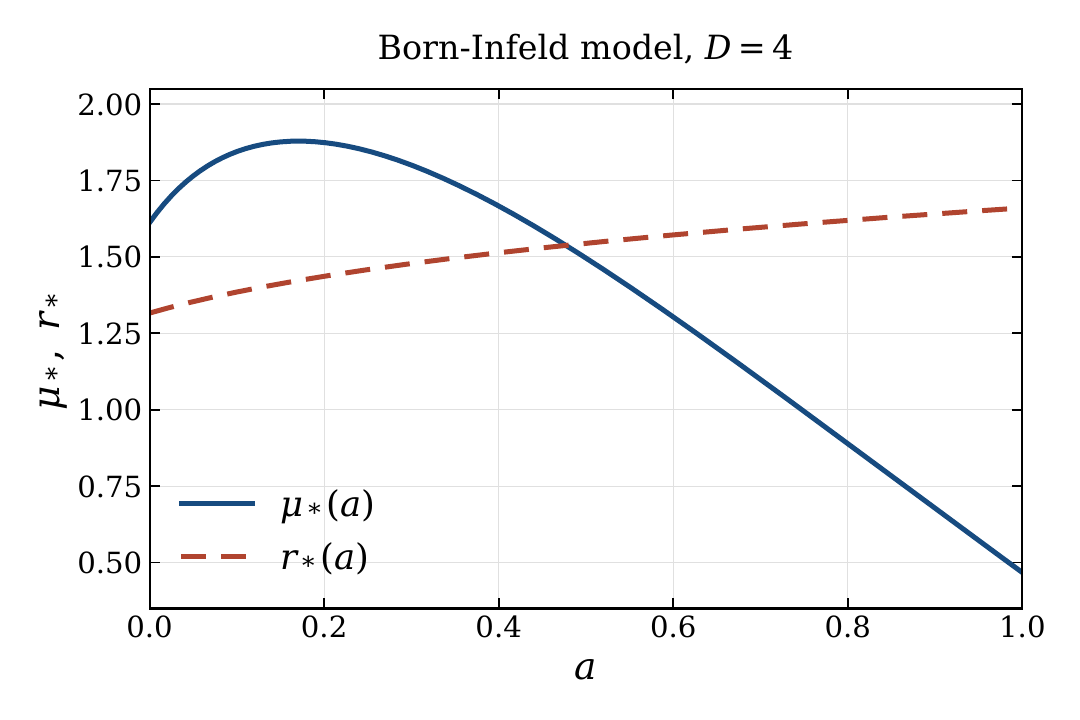}\\[0pt]
\caption{$\mu_*$ and $r_*$ as a function of $a$ for Born-Infeld-type QTG model.}
    \label{F2}
\end{figure}

\begin{figure}[!htb]%
    \centering \includegraphics[width=0.4\textwidth]{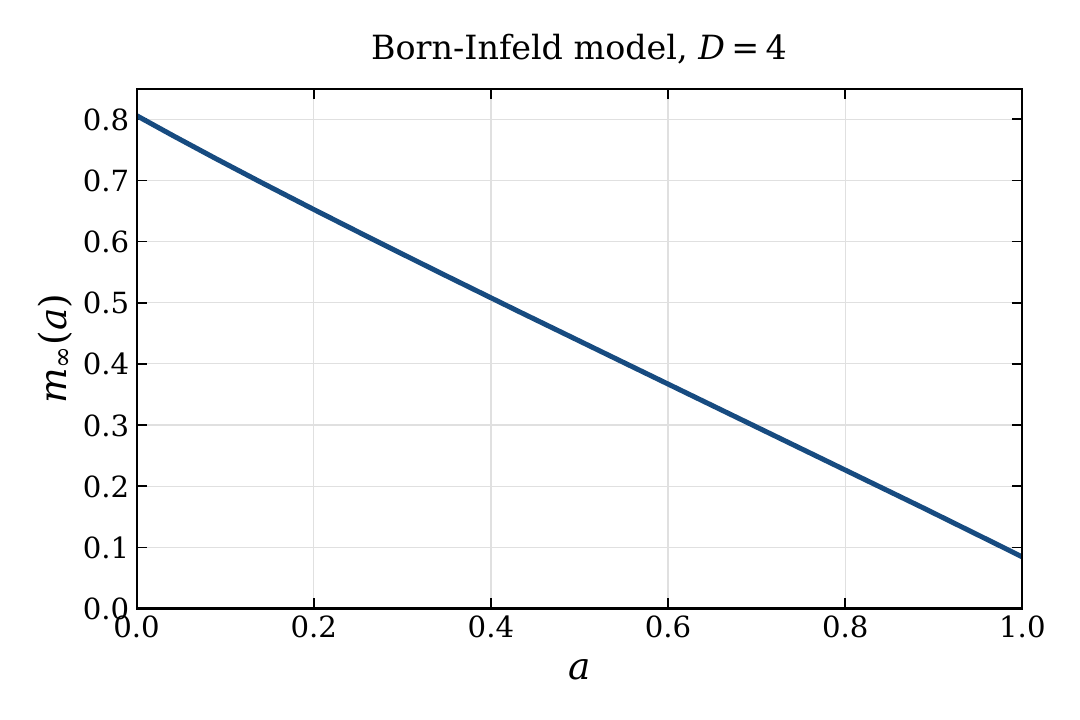}\\[0pt]
\caption{Mass at infinity $m_{\infty}$ of an extremal minimons of the Born-Infeld-type QTG model as a function of $a$.}
    \label{F2}
\end{figure}

In these samples the central parameter is nonmonotonic in $a$,
whereas the asymptotic mass decreases. Reducing $\epsilon$ changes
$\mu_*$ and $m_\infty$ only slightly but increases $\nu$ substantially.
These finite-$\epsilon$ data do not establish a universal power law
for $\nu$ or a global existence result for all $a$.

\vspace{0.5cm}

{ }

\begin{figure}[!htb]%
    \centering \includegraphics[width=0.4\textwidth]{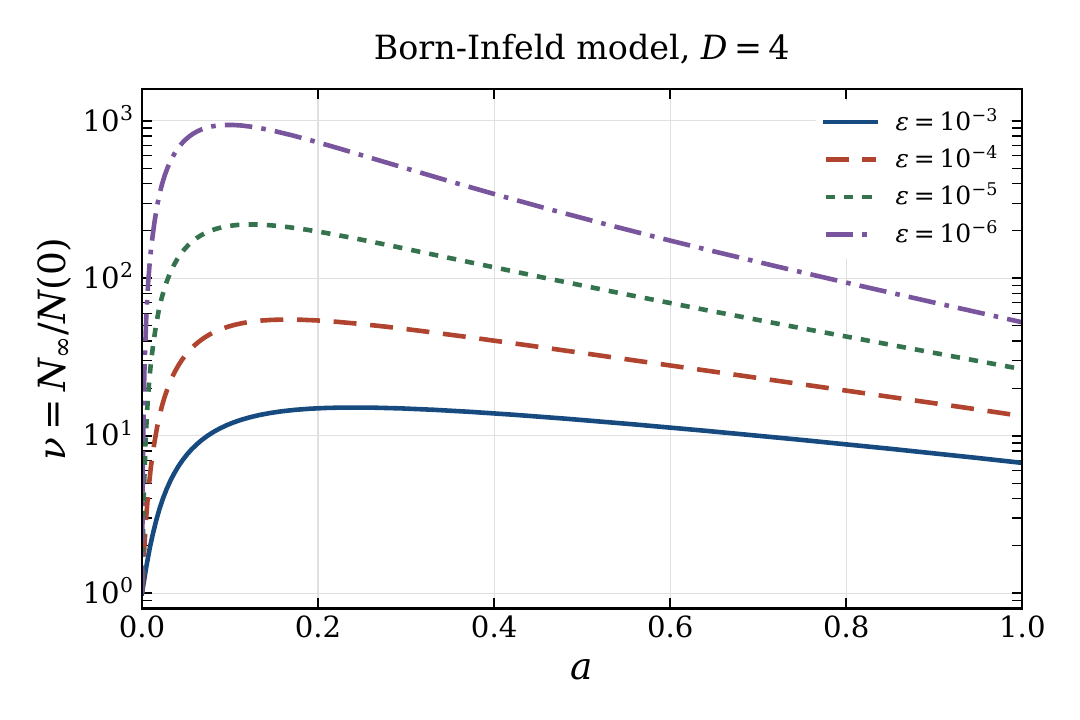}\\[0pt]
\caption{$\nu=N_{\infty}/N_0$ as a function of $a$ for different choice of a small parameter $\epsilon$ for Born-Infeld QTG model.}
    \label{F6}
\end{figure}

\vspace{0.5cm}

{ }

\begin{figure}[!htb]%
    \centering \includegraphics[width=0.4\textwidth]{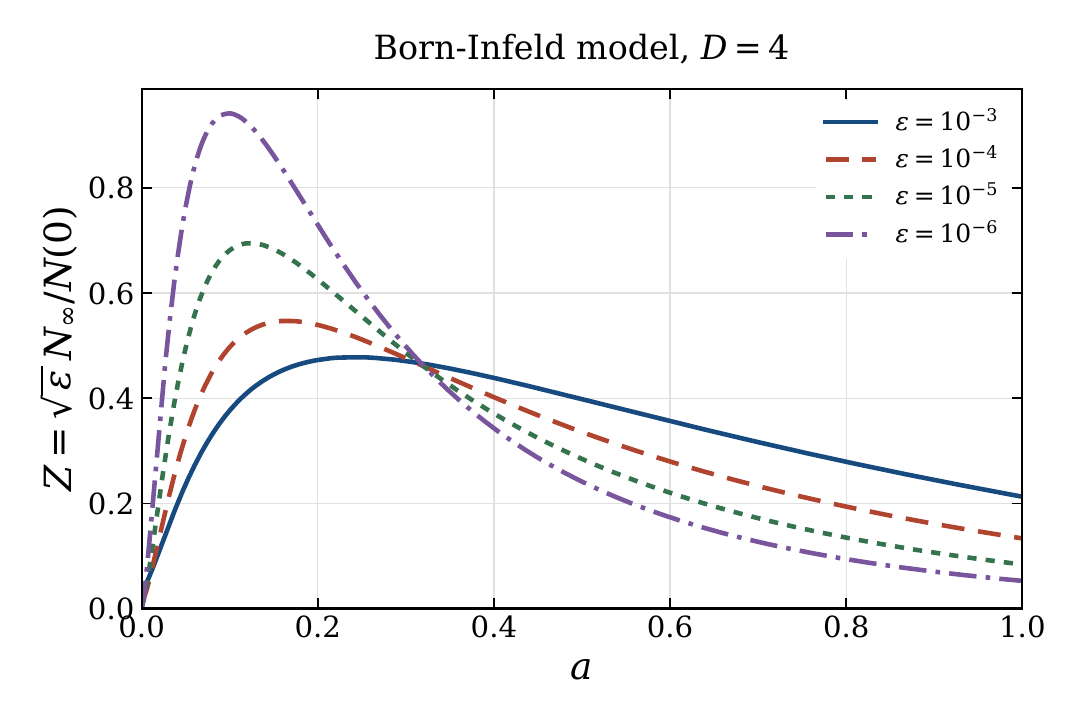}\\[0pt]
\caption{$Z=\sqrt{\epsilon}N_{\infty}/N_0$ as a function of $a$ for different choice of a small parameter $\epsilon$ for Born-Infeld QTG model.}
    \label{F7}
\end{figure}

\section{QTG extremons: Near-horizon geometry}

We refer to the critical limiting configurations of QTG minimons as QTG
extremons. In this section, we derive their constant-radius throat geometry
under the assumption that the near-critical solutions admit a smooth local
limit in proper distance, with finite two-dimensional curvature, finite
$h$, $h_p$, and $h_{pp}$, and $r_*\to r_*>0$. This construction determines
a candidate near-horizon geometry. Whether this geometry is attained by
a given family of global solutions, and whether the corresponding limiting
horizon is regular, require separate analyses. The numerical plots presented
above do not resolve these questions.

\subsection{Proper-distance derivation}
In a region with $f>0$, introduce
\begin{equation}
\begin{split}
 &u=\sqrt{f}=\frac{\dd r}{\dd x},\hh
 \mathcal A=N u, \\
 & k=\frac{\mathcal A'}{\mathcal A}, \hh Q=\varphi'=-2k\, .
 \end{split}
\end{equation}
where primes henceforth denote $\dd/\dd x$. Then
\begin{equation}
 ds^2=-\mathcal A(x)^2\dd t^2+\dd x^2+r(x)^2\dd\Omega_{D-2}^2 \, .
\end{equation}
Transformation of the source equations gives
\begin{equation}\label{proper-eqs}
 \begin{aligned}
 2\mathcal W r''-J+4ak^2+8ak'&=0,\\
 2\mathcal W r'k-J+4ak^2&=0.
 \end{aligned}
\end{equation}
These imply
\begin{equation}
 \mathcal W(r''-r'k)+4ak'=0.
\end{equation}
The radius equation, retained before taking a constant-radius limit, is
\begin{equation}\label{radius-eq}
 2\mathcal W(k'+k^2)
 +2(\mathcal W_{,r}r'+\mathcal W_{,u}r'')k
 +2\mathcal W_{,r}r''-J_{,r}=0.
\end{equation}
The commas denote partial derivatives of functions of $(r,u)$, with
$p=(1-u^2)/r^2$. This equation follows by differentiation of the
constraint where $r'\ne0$, or by variation of the radius in the reduced
action. It must not be discarded when $r'\to0$.

At the minimum, $r'=\sqrt\epsilon$ and $r''=f_r/2=0$.
In a smooth constant-radius limit these derivatives vanish, giving
\begin{equation}\label{throat-algebra}
 J_0=4ak_0^2,\qquad k_0'=0,\qquad
 2\mathcal W_0 k_0^2=J_{,r}(r_*,0).
\end{equation}
For $\mathcal W_0\ne0$ and $k_0^2>0$, define
\begin{equation}\label{general-radius}
 L^{-2}=k_0^2=\frac{J_{,r}(r_*,0)}{2\mathcal W_0},\qquad
 J_0\mathcal W_0=2aJ_{,r}(r_*,0).
\end{equation}
These are the general QTG throat-radius equations. If their positivity
or nondegeneracy assumptions fail, this particular AdS throat
construction does not apply.

Choosing the orientation with $k_0=1/\ell_*$ and rescaling time gives
\begin{equation}\label{proper-throat}
ds^2_{\rm NH}=-e^{2x/\ell_*}\dd T^2+\dd x^2
 +r_*^2\dd\Omega_{D-2}^2.
\end{equation}
To obtain this local metric from the family one uses
$T=\mathcal A(x_*)t$ and places $x_*=0$ at the minimum; this time
normalization is local and can become singular relative to time at
infinity. The opposite sign of $k_0$ is obtained by reversing $x$.
With $z=e^{x/\ell_*}>0$ and $\tau=T/\ell_*$,
\begin{equation}\label{ads-throat}
ds^2_{\rm NH}=\ell_*^2\left(-z^2\dd\tau^2+
\frac{\dd z^2} {z^2}\right)+r_*^2\dd\Omega_{D-2}^2.
\end{equation}
Thus the local limiting geometry is $\mathrm{AdS}_2(\ell_*)\times
S^{D-2}(r_*)$. Its two-dimensional curvature is $R=-2/\ell_*^2$,
and $z=0$ is the degenerate Poincare horizon. The physical metric is
$\ell^2$ times this dimensionless metric.

The auxiliary field and its proper-distance gradient are
\begin{equation}
 \varphi=\varphi_0-\frac{2x}{\ell_*}=\varphi_0-2\ln z,\qquad
 Q=-\frac{2}{\ell_*}.
\end{equation}
Although $Q$ is constant and nonzero, the integration constant is
\begin{equation}
 K=\mathcal A Q+2\mathcal A'=0.
\end{equation}
The lapse $\mathcal A$ is exponential, not constant. Hence this throat
does not contradict the nonexistence of a constant-lapse,
constant-radius solution with $aK\ne0$.
On the exact product geometry the physical two-dimensional Polyakov
tensor is
\begin{equation}
 T^{\rm P}_{ab}=-\frac{\hbar C}{24\pi\ell^2\ell_*^2}\gamma_{ab},\qquad
 \rho_{\rm P}=\frac{\hbar C}{24\pi\ell^2\ell_*^2},\qquad
 P_{\rm P}=-\rho_{\rm P},
\end{equation}
where the tensor coefficient is understood relative to the physical
metric. It has the required trace anomaly and is regular on the
extended exact AdS factor, even though the auxiliary field diverges
at the Poincare horizon.
To illustrate the above relations, we consider two specific four-dimensional QTG models: one of Hayward type and the other of Born–Infeld type.

\subsection{$4D$ Hayward-type QTG model}

\begin{figure}[!htb]%
    \centering \includegraphics[width=0.4\textwidth]{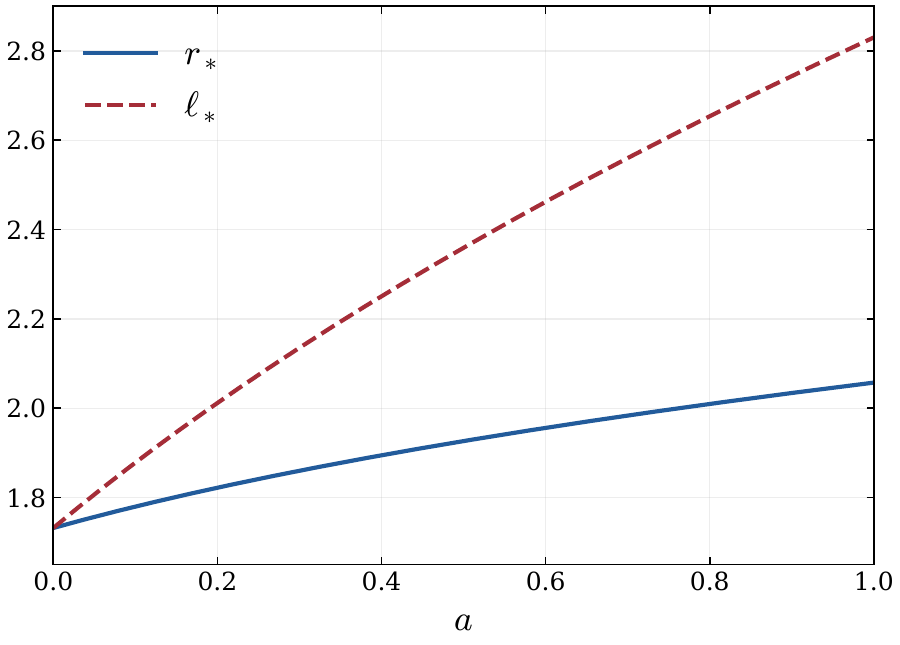}\\[0pt]
\caption{$r_*$ and $\ell_*$ as functions of $a$ for a QTG extremon in the Hayward-type QTG model with $D=4$.}
    \label{F9}
\end{figure}

For the model in the source,
\begin{equation}
 h=\frac{p}{1-p},\, h_p=\frac1{(1-p)^2},\,
 h_{pp}=\dfrac{2}{(1-p)^2}\, .
\end{equation}
Put $y=r_*^2$. On the branch $p_*<1$, one has $y>1$. At $u=0$,
\begin{equation}
\begin{split}
 &\mathcal W_0=\frac{y^{5/2}}{(y-1)^2},\quad
 J_0=\frac{y(y-3)}{(y-1)^2},\\
& J_{,r}(r_*,0)=\frac{2\sqrt y(y+3)}{(y-1)^3}.
\end{split}
\end{equation}
Substitution into the throat equations gives
\begin{equation}
\begin{split}
\label{hayward-radii}
& a(y)=\frac{y^3(y-3)}{4(y+3)(y-1)},\\
 &r_*^2=y,\qquad \ell_*^2=\frac{y^2(y-1)}{y+3}.
\end{split}
\end{equation}
For $a>0$, the branch connected to the vacuum extremal solution has
$y>3$. It is unique: the function
is strictly increasing from zero to infinity on $y>3$.
Relations \eqref{hayward-radii} give parametric representation of the  parameters $r_*$ and $\ell_*$ for the solution \eqref{ads-throat} as functions of $a$. These functions are presented in the figure~\ref{F9}

\vspace{0.5cm}

{ }

The near-horizon metric is therefore
\begin{equation}
ds^2_{\rm NH}=
 \frac{y^2(y-1)}{y+3}
 \left(-z^2\dd\tau^2+\frac{\dd z^2}{z^2}\right) +y\dd\Omega_2^2.
\end{equation}
In particular, the two curvature radii are unequal when $a>0$.
The small-$a$ expansion is
\begin{equation}
 r_*^2=3+\frac{16}{9}a+O(a^2),\qquad
 \ell_*^2=3+\frac{16}{3}a+O(a^2).
\end{equation}

\subsection{4D Born--Infeld-type QTG model}
\label{subsec:bi-extremon4}

\begin{figure}[!htb]%
    \centering \includegraphics[width=0.4\textwidth]{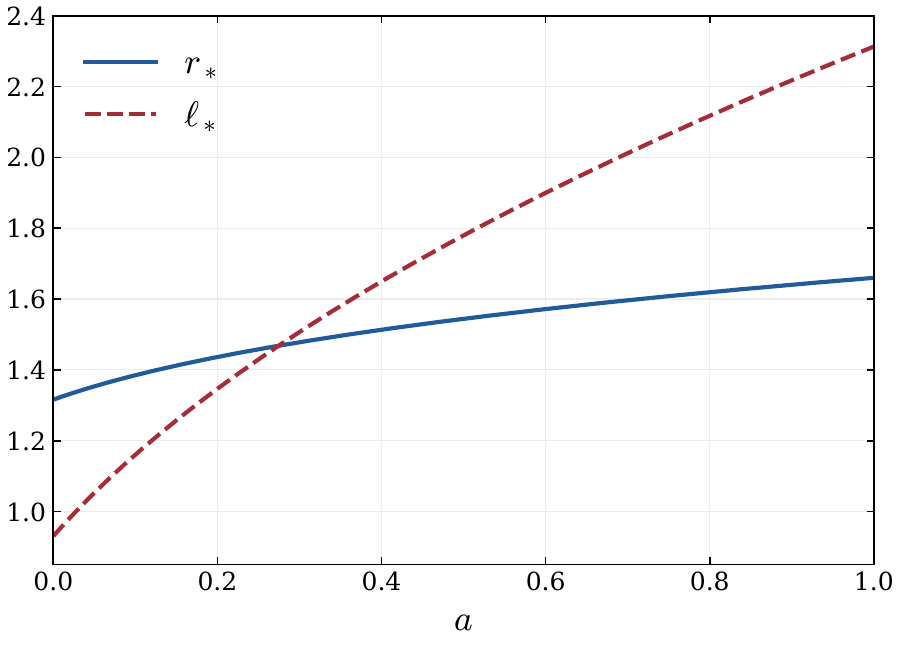}\\[0pt]
\caption{$r_*$ and $\ell_*$ as functions of $a$ for a QTG extremon in the Born-Infeld-type QTG model with $D=4$.}
    \label{F10}
\end{figure}

For the Born--Infeld-type model,
\begin{equation}
\begin{split}
 h&=\frac{p}{\sqrt{1-p^2}},\qquad
 h_p=\frac{1}{(1-p^2)^{3/2}},\\
 h_{pp}&=\frac{3p}{(1-p^2)^{5/2}}, .
\end{split}
\end{equation}
We use the same smooth constant-radius limit as above and put
$y=r_*^2$. Since $p_*=1/y$, the positive-curvature Born--Infeld
branch requires $y>1$. At $u=0$, the coefficients in the throat
equations are
\begin{equation}
\begin{split}
 \mathcal W_0&=\frac{y^{7/2}}{(y^2-1)^{3/2}},\qquad
 J_0=\frac{y(y^2-3)}{(y^2-1)^{3/2}},\\
 J_{,r}(r_*,0)&=\frac{6\sqrt y\,(y^2+1)}{(y^2-1)^{5/2}}.
\end{split}
\end{equation}
Here $J_{,r}$ is evaluated at fixed $u$. Substitution into
$J_0\mathcal W_0=2aJ_{,r}$ and $\ell_*^2=2\mathcal W_0/J_{,r}$ gives
\begin{equation}
\begin{split}
&a=\dfrac{ y^4(y^2-3)}{12(y^2+1)\sqrt{y^2-1}}, ,\\
&r_*^2=y,\qquad
 \ell_*^2=\frac{y^3(y^2-1)}{3(y^2+1)}.
\end{split}
\label{eq:bi-extremon-radii}
\end{equation}
For $a>0$, the branch connected to the vacuum extremal solution
has $y>\sqrt3$. It is unique: the function
$a$
is strictly increasing from zero to infinity on this interval.
Relations \eqref{eq:bi-extremon-radii} give parametric representation of the  parameters $r_*$ and $\ell_*$ for the solution \eqref{ads-throat} as functions of $a$. These functions are presented in the figure~\ref{F10}

\vspace{0.5cm}

{ }

The small-$a$ expansion is
\begin{equation}
\begin{split}
 r_*^2&=\sqrt3+\frac{8\sqrt6}{9}a+O(a^2),\\
 \ell_*^2&=\frac{\sqrt3}{2}+2\sqrt6\,a+O(a^2).
\end{split}
\label{eq:bi-extremon-small-a}
\end{equation}
Thus, already in the vacuum limit, the two curvature radii differ:
$\ell_*^2=r_*^2/2$ at $a=0$. These local throat relations do not
determine the central critical parameter or the asymptotic mass.
Their realization as the limit of a global family, and the
regularity of that family's limiting horizon, require a separate
analysis.

\section{Discussion}

We have studied the regular horizonless branch of static,
spherically symmetric solutions in QTG and its modification by vacuum
polarization. In the absence of quantum matter, these configurations
exist below the mass threshold at which the inner and outer black
hole horizons merge. We refer to them as QTG minimons. The Hayward-type
and Born--Infeld-type models provide two explicit examples with a
bounded-curvature, de Sitter-like central region.

The Polyakov action allows us to include the backreaction of
two-dimensional conformal fields while retaining a tractable system
of equations. We considered a static auxiliary field and imposed
$K=0$, corresponding to an asymptotically empty state. For a general
characteristic function $h(p)$, the field equations reduce to a
first-order system for $H$, $N$, and $q$, supplemented by an algebraic
constraint. The central expansions provide regular limiting data
for numerical integration and show that the de Sitter-like core
persists in both models. The approach to the core is model dependent:
the first correction to $f=1-r^2$ occurs at order $r^{D+1}$ for
Hayward and at order $r^{2D}$ for Born--Infeld. These expansions
establish finite central curvature, without implying smoothness to
arbitrary order in Cartesian coordinates.

The central parameter $H_0$ and the mass measured at infinity must be
distinguished once vacuum polarization is included. In four dimensions,
$m_\infty=H_\infty/2$, whereas $H_0/2$ differs from this quantity by an
integrated matter contribution. In the range of couplings studied,
the near-critical asymptotic mass decreases approximately linearly
with $a$ in both models, although the critical central parameter
$H_{0,\mathrm{crit}}(a)$ is nonmonotonic. The approximate linearity of
the mass is a numerical property of these solutions, rather than an
exact relation established by the field equations.

A more pronounced effect appears in the lapse ratio
$\nu=N_\infty/N(0)$. This ratio is invariant under a constant rescaling
of time and determines the frequency shift of a photon arriving at
the center from infinity: $\omega_0=\nu\omega_\infty$. At $a=0$, the
lapse is constant and $\nu=1$. For fixed positive $a$, the numerical
solutions show a strong increase in $\nu$ as the minimum
$\epsilon=f_{\min}$ decreases. The rapid variation of the lapse is
concentrated near this minimum. Thus, a small quantum coupling does
not guarantee a small correction throughout the near-critical
geometry.

The rescaled quantity $Z=\sqrt{\epsilon}\,\nu$ remains of order unity
over substantial parts of the parameter range displayed, but its
curves retain a dependence on $\epsilon$. Our results therefore do
not establish a universal law $\nu\propto\epsilon^{-1/2}$. In
particular, the local growth of $N'/N$ at the minimum does not by
itself determine the integrated lapse ratio. A more detailed analysis
of the transition region and its matching to the central and
asymptotic regions is needed to determine the limiting behavior of
$\nu$.

We call the critical limiting configurations QTG extremons. Under
the assumption of a smooth constant-radius limit in proper distance,
the field equations determine a local product geometry
$\mathrm{AdS}_2(\ell_*)\times S^{D-2}(r_*)$. In four dimensions, both models
admit a unique candidate throat radius on the branch connected to
the vacuum extremal solution. The two curvature radii depend on the
model and on $a$; they need not coincide. On the exact product
geometry, the Polyakov stress-energy tensor is regular and
proportional to the two-dimensional metric, despite the logarithmic
behavior of the auxiliary field at the Poincare horizon. This local
result does not establish the regularity of the limiting horizon of
a global minimon family. The relation between the local throat time
and the time normalized at infinity can become singular in the
critical limit.

Several questions remain. The dynamical stability of minimons and
their formation from regular initial data have not been addressed.
Likewise, the static solutions alone do not determine whether
minimons or extremons can arise as endpoints of black hole
evaporation. These questions require time-dependent evolution and
appropriate quantum-state conditions. It is also necessary to test
how the results change when the matter effective action includes
the dilaton-dependent terms associated with spherical reduction of
higher-dimensional fields. The Polyakov model captures a specific
vacuum-polarization mechanism, but is not the full effective action
of four-dimensional quantum matter. Within this model, the main
conclusion is that bounded central curvature can coexist with a
large vacuum-induced modification of the lapse near the threshold
for horizon formation.



%

\end{document}